\documentclass[useAMS,usenatbib]{mn2e}

\usepackage{amssymb,graphicx}

\title[Hadronic gamma--rays from RX~J1713]
{Hadronic gamma--rays from RX~J1713.7-3946?}
\author[S. Gabici and F. A. Aharonian]{S. Gabici$^{1}$\thanks{E-mail:
stefano.gabici@apc.univ-paris7.fr} and F. A.
Aharonian$^{2,3}$\\
$^{1}$APC, Univ Paris Diderot, CNRS/IN2P3, CEA/Irfu, Obs de Paris, Sorbonne Paris Cit\'e, France\\
$^{2}$Dublin Institute for Advanced Studies, 31 Fitzwilliam Place, Dublin 2, Ireland \\
$^{3}$Max-Planck-Institut f\"ur Kernphysik, Postfach 103980, D-69029 Heidelberg, GERMANY }

\begin{document}

\date{Accepted 1988 December 15. Received 1988 December 14; in original form 1988 October 11}

\pagerange{\pageref{firstpage}--\pageref{lastpage}} \pubyear{2002}

\maketitle

\label{firstpage}

\begin{abstract}
RX~J1713.7-3946 is a key object to check the supernova remnant paradigm of the origin of Galactic  cosmic rays. 
While the origin of its gamma--ray emission (hadronic versus leptonic) is still debated, the hard spectrum at GeV energies reported by the {\it Fermi} collaboration is generally interpreted as a strong argument in favor of a leptonic scenario.
On the contrary, we show that hadronic interactions can naturally explain the gamma--ray spectrum if gas clumps are present in the supernova remnant shell. The absence of thermal X--rays from the remnant fits well within this scenario.
\end{abstract}

\begin{keywords}
cosmic rays -- ISM: supernova remnants -- gamma rays: ISM.
\end{keywords}

\section{Introduction}

SuperNova Remnants (SNRs) are believed to be the sources of Galactic Cosmic Rays (CRs). 
This hypothesis, still unproven, received further support after the detection of several SNRs in TeV gamma rays. 
Such emission was indeed expected, as the result of the decay of neutral pions produced in the interactions between the accelerated CRs and the ambient gas \citep{dav}. However, 
the relative contribution of protons and electron (via inverse Compton scattering) to the gamma--ray emission remains uncertain, being very sensitive to model parameters such as gas density and magnetic field strength, which are often unknown \citep[e.g.][]{pierre}. In most cases, this prevents the unambiguous identification of CR protons. 


RX~J1713.7-3946 is the best studied TeV SNR \citep{RXJHESS1}. After an initial debate on the hadronic versus leptonic origin of its gamma--ray emission \citep{berezhko,giovanni, vladfelix}, it was detected by {\it Fermi} \citep{fermi}. The hard spectrum revealed in the GeV domain seemed incompatible with that expected from shock accelerated protons and leptonic models became favored over hadronic ones \citep{don,yuan,finke}. Moreover, 
the large gas density required by hadronic models would imply an intense thermal X--ray line emission \citep{katz,don}, which is not observed \citep{tanaka}.
However, the leptonic interpretation is not exempt from problems. In fact, it is known that one--zone leptonic models, when normalized to fit the TeV data, fail to reproduce the GeV flux \citep[e.g.][]{felixSNRs}. 

As suggested by \citet{vladfelix}, the interpretation of the gamma--ray observations changes dramatically if the SNR expands in a clumpy medium. This is indeed expected if the SNR progenitor is a massive star in a molecular cloud. The densest clumps can survive, unshocked, the shock passage, and remain inside the SNR \citep{inoue2012}. Then, the CRs accelerated at the SNR diffusively penetrate the clumps. Since the diffusion coefficient is an increasing function of particle energy, higher energy particles penetrate more effectively, and the spectrum of the CRs inside the clumps might well be significantly harder than the one accelerated at the shock \citep{atoyan,gabici07}. Thus, if clumps  make the dominant contribution to the mass in the SNR, hadronic emission can naturally explain the hard spectrum observed by {\it Fermi}. Moreover, the fact that clumps remain unshocked implies that most of the gas in the shell is at low temperature. This would explain the lack of X--ray lines in the spectrum.

Here, we develop a model to describe this scenario, and demonstrate that it provides an excellent fit to data. 

\section{Dynamical evolution of the SNR}


We assume here that the progenitor of the SNR RX~J1713.7-3946 is a massive star embedded in a molecular cloud. This scenario is supported by a number of observational evidences \citep[][]{slane99,moriguchi2005,fukui2012,maxted2012}. The stellar wind from the progenitor strongly affects the properties of the local environment by inflating a large cavity of hot and rarefied gas \citep{weaver77}. Molecular clouds are highly inhomogeneous, clumpy structures, and the densest clumps survive the wind and remain intact in the cavity \citep{inoue2012}. In this scenario, the SNR shock propagates in the low density gas, which occupies most of the volume, and interacts with dense clumps characterized by a very small volume filling factor.

The temporal evolution of the SNR shock radius $R_s$ and velocity $u_s = {\rm d}R_s/{\rm d}t$ can be computed by adopting the thin--shell approximation \citep[e.g.][]{ostrikermckeereview}. The equation of momentum conservation then reads:
\begin{equation}
\label{eq:momentum}
\frac{{\rm d}(Mu)}{{\rm d}t} = 4 \pi ~ R_s^2 ~ P_{th}
\end{equation}
where $u = (3/4) u_s$ and $P_{th}$ are the gas velocity and thermal pressure behind the shock, respectively. The pressure of the ambient medium is neglected in Eq.~\ref{eq:momentum} because SNR shocks are strong. The pressure $P_{th}$ can be derived from the conservation of the explosion energy $E_{tot} = 10^{51} E_{51}$~erg:
\begin{equation}
\label{eq:energy}
E_{tot} = E_{th}+(1/2) M u^2
\end{equation}
where $E_{th}$ is the thermal energy inside the SNR, and $M$ is the mass of the gas in the SNR shell, which is the sum of the mass of the supernova ejecta $M_{ej}$ and of the ambient gas  of density $\varrho$ swept up by the shock: $M_{sw} = 4 \pi \int_0^{R_s} {\rm d}r ~ r^2 \varrho(r)$. The swept up mass is obtained by recalling that the SNR shock expands first in the progenitor's wind, with density profile $\varrho \propto r^{-2}$, and then in the hot and tenuous gas that fills the cavity, of Hydrogen density $n_h = 10^{-2} n_{h,-2}$~cm$^{-3}$ and temperature $T_h = 10^6 ~ T_{h,6}$~K. The density profile of the wind is $n_w = \dot{M}/(4 \pi m_a u_w r^2)$, where $\dot{M} = 10^{-5} \dot{M}_{-5} M_{\odot}/$yr is the wind mass loss rate, $u_w = 10^6 u_{w,6}$~cm/s is the wind speed, and $m_a = 1.4 ~ m_p$ is the mean mass of interstellar nuclei per Hydrogen nucleus for $10$\% Helium abundance. The position of the wind termination shock is determined by imposing equilibrium between the wind ram pressure $n_w m_a u_w^2$ and the thermal pressure in the cavity $2.3 n_h k_B T_h$, the factor 2.3 indicating the number of particles per Hydrogen atom.

Eqns.~\ref{eq:momentum} and \ref{eq:energy} can be solved for the case of RX~J1713.7-3946. To reproduce the measured radius of the shock $R_{s,0} \approx 10$~pc at the SNR age of $\approx 1620$~yr \citep{wang} the following parameters have been adopted: 
$M_{ej}  = 2.8 ~ M_{\odot}$, $n_{h,-2} = 2$, and $E_{51} = \dot{M}_{-5} = u_{w,6} = T_{h,6} = 1$
For such a choice of the parameters, the mass of the swept up gas is $M_{sw} \sim 3.8 ~M_{\odot}$, which is comparable to the mass of the ejecta. This implies that the SNR is still in the transition between the ejecta--dominated and the Sedov phase, and that the shock speed is still quite large. Our calculations give $u_{s,0} \sim 4.4 \times 10^3$~km/s, in agreement with the observational constrain given in \citet{yas}.

The clumps that survive the stellar wind and remain embedded in the cavity are characterized by a sub--parsec size, $L_c = 0.1~ L_{c,-1}$~pc, and a large density, $n_c \gtrsim 10^3 n_{c,3}$~cm$^{-3}$ (see the simulations by \citealt{inoue2012} and the observations by \citealt{fukui2012} and \citealt{sano2013}). 

The main parameter that regulates the interaction between a clump and the shock is the density contrast between the clump and the diffuse medium, $\chi = n_c/n_h = 10^5 ~ n_{c,3}/n_{h,-2}$ \citep{klein94}. When the SNR shock encounters a clump, it drives a shock into it, with a velocity $u_c \approx u_s/\chi^{1/2}$. The time it takes the clump to be shocked, $t_{cc} \equiv L_c/u_c$, is called {\it cloud crushing time}. \citet{klein94} showed that the typical time scale for the development of Kelvin--Helmholtz and Rayleigh--Taylor plasma instabilities that might disrupt the clump are of the same order of $t_{cc}$.
The value of the cloud crushing time for RX~J1713.7-3946 is $t_{cc} \approx 7 \times 10^3 L_{c,-1} (n_{c,3}/n_{h,-2})^{1/2} (u_s/u_{s,0})^{-1}$~yr. This time is significantly longer than the SNR age, which means that the clumps survive against plasma instabilities. This fact has been confirmed by numerical simulations, which show that shocks are stalled into dense clumps, which survive, unshocked, in the SNR interior \citep{inoue2012}. 
Clump evaporation due to thermal conduction is neglected here, being strongly suppressed by the turbulent magnetic field downstream of the SNR shock \citep{chandran}.

Another consequence of the long cloud crushing time is the fact that the clumps remain virtually at rest (in the lab frame) after the shock passage. The large difference between the velocity of the clump and that of the shocked medium generates a velocity shear which is in turn responsible for the amplification of the magnetic field in a boundary layer around the clump. The magnetic field grows very quickly (tens of years) to large values of the order of $\gtrsim 100~\mu$G, with peak values up to $\sim 1$~milliGauss, in a transition layer of size $L_{tr} \sim 0.05$~pc \citep{inoue2012}.
The magnetic field in the whole SNR shell is also amplified to tens of microGauss due to the vorticity induced downstream by the deformation of the shock surface. Such deformations can be produced by the interaction of the shock with dense clumps \citep{giacalone}. Another amplification mechanism has to be considered if the SNR shock accelerates effectively CRs, i.e. the CR current driven instability, which predicts a magnetic field pressure downstream of the shock at the percent level of the shock ram pressure, $B_d \sim 3 \times 10^{-5} n_{h,-2}^{1/2} (u_s/u_{s,0}) ~ \mu$G \citep{bell2013}, which is of the same order of the field amplified by the shock--induced vorticity.
 
In the following we assume that the gas is clumpy inside the SNR shell, with a magnetic field of few tens of microGauss in the diffuse gas, and of $\gtrsim 100~\mu$G in a thin transition region surrounding the clumps.

\section{Gamma--ray emission}

The SNR shock is expected to accelerate CRs. For a strong shock, the test--particle prediction from shock acceleration theory gives an universal power law spectrum of accelerated particles $Q_{CR}(E) \propto E^{-\alpha}$ with $\alpha = 2$ \citep[e.g.][]{lukereview}. The particle spectra inferred from gamma--ray observations of SNRs are somewhat steeper than that, and several modifications to the shock acceleration mechanism have been proposed to explain the discrepancy \citep[][]{vladimirdrift,damiano}. In the following we assume $\alpha = 2.2$.

The CR acceleration efficiency $\eta$ is defined as the fraction of the kinetic energy flux across the shock which is converted into CRs:
\begin{equation}
q_{CR} = \int_{m_p c^2}^{E_{max}} {\rm d}E ~ E ~ Q_{CR}(E) =  \eta ~ \frac{1}{2} \varrho u_s^3 ~ (4 \pi R_s^2) ~ .
\end{equation}
The maximum energy of accelerated particles $E_{max}$ is computed by equating the CR diffusion length $D_B/u_s$,  $D_B$ being the Bohm diffusion coefficient, to some fraction $\chi$ of the shock radius \citep[e.g.][]{meescape}. This gives $E_{max} \sim 200 (R_s/R_{s,0}) (u_s/u_{s,0}) (\chi/0.05) B_{up,-5}$~TeV, where $B_{up} = 10 ~ B_{up,-5} \mu$G is the magnetic field upstream of the shock.

Once accelerated, CRs are advected downstream of the shock where they suffer adiabatic energy losses due to the expansion of the SNR at a rate $\dot{E} = E (u_s/R_s)$. The time evolution of the total number of CRs inside the SNR shell $N_{CR}(E)$ is described by the equation:
\begin{equation}
\label{eq:shell}
\frac{\partial N_{CR}(E)}{\partial t} = \frac{\partial}{\partial E} \left[ \dot{E} N_{CR}(E) \right] + Q_{CR}(E) ~ .
\end{equation}

Consider now a clump entering the SNR shock at a time $t_c$. Once downstream of the shock, the clump is bombarded by the CRs accelerated at the SNR shock and accumulated in the SNR shell. The diffusion of CRs in the highly turbulent region that surrounds the clump is expected to occur at the Bohm rate. Thus, the time needed for a CR to diffusively penetrate into the clump is $\tau_d \approx L_{tr}^2/6 D_B$ which gives:
\begin{equation}
\label{eq:difftime}
\tau_d \approx 4 \times 10^2 ~ L_{tr,-1.3}^2 B_{-4} E_{12}^{-1} ~ {\rm yr}
\end{equation}
where $B = 100 ~ B_{-4} ~ \mu$G is the magnetic field in the turbulent layer, $L_{tr} = 0.05 ~ L_{tr,-1.3}$~pc its thickness, and $E = E_{12}$~TeV the particle energy. For a given clump, the minimum energy of the particles that can penetrate is given by the equation $\tau_d = t_{age}-t_c$. A significantly faster CR diffusion is expected outside of the transition region, both inside the clump, where ion-neutral friction is expected to heavily damp magnetic turbulence, and in the SNR shell, where the magnetic field strength and turbulent level are significantly smaller.

The equation that regulates the time evolution of the total number of CRs inside a clump $N_{cl}(E)$ is then: 
\begin{equation}
\label{eq:clump}
\frac{\partial N_{cl}(E)}{\partial t} = \frac{(V_{cl}/V_{sh})N_{CR}(E) - N_{cl}(E)}{\tau_d}
\end{equation}
where $V_{cl} = (4 \pi /3) L_c^3$ and $V_{sh}$ are the volumes of the clump and of the SNR shell, respectively. The total volume of the clumps is taken to be much smaller than $V_{sh}$, to insure the validity of Eq.~\ref{eq:shell}. Moreover, $V_{cl}$ is assumed to be constant in time (i.e. no CR adiabatic energy losses) and proton--proton interaction energy losses are neglected since they operate on a time $t_{pp} \sim 5 \times 10^4 n_{c,3}^{-1}$~yr, longer than the age of the SNR. Finally, the volume $V_{sh}$ filled by CRs is taken to be the shell encompassed between the SNR forward shock and the contact discontinuity. The exact position of the contact discontinuity depends on several physical parameters \citep[e.g.][]{salvatoreCD}, and is typically of the order of $\approx 0.9 R_s$.


 
\begin{figure}
\centering
\includegraphics[width=.39\textwidth]{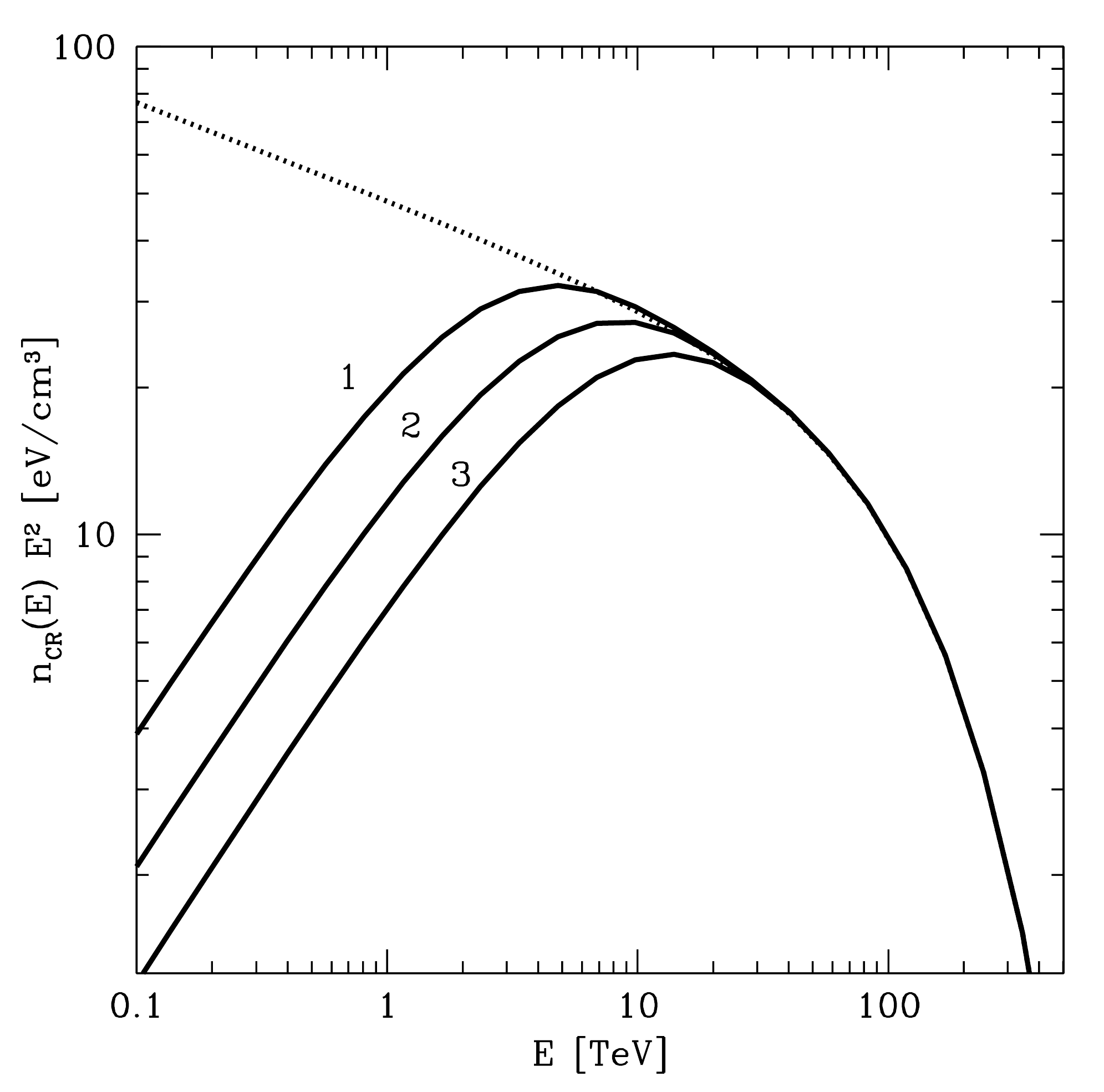}
\caption{Spectrum of CRs in the SNR shell (dotted line) and inside a clump that entered the shock at $t_c = 1400$, $1500$, and $1550~yr$ (solid line 1, 2, and 3 respectively).}
\label{fig1}
\end{figure}

The dotted line in Fig.~\ref{fig1} represents the current CR density in the SNR shell as a function of the particle energy. It has been computed from Eq.~\ref{eq:shell} after assuming a CR acceleration efficiency of $\eta = 0.1$ and a magnetic field in the turbulent layer of $B_{-4} = 1.2$. An exponential cutoff at $E_{max} = 150$~TeV  has been multiplied to the solution of Eq.~\ref{eq:shell} to mimic the escape of the highest energy CRs  from the shock. The CR density inside clumps is derived from Eq.~\ref{eq:clump} and plotted with solid lines. Lines 1, 2, and 3 refer to a clump that entered the SNR shock 1400, 1500, and 1550 yr after the supernova explosion, respectively. Clumps that entered the SNR at $t_c \approx 1400$~yr are encountering now the contact discontinuity. Clumps that entered the SNR earlier either are disrupted by plasma instabilities at the contact discontinuity or, if they survive, enter a region characterized by a low density of CRs and quickly become themselves devoided of CRs due to their diffusive escape.

The spectrum of CRs inside the clumps has a characteristic peak at energies of $\approx 10$~TeV. At energies larger than that of the peak, the spectra of the CRs in the clumps and in the SNR shell coincide. This is because at large energies diffusion becomes important over times smaller than the residence time of clumps in the shell, allowing for a rapid equilibration of CR densities. On the other hand, CRs with energies smaller than that of the peak diffuse too slowly to effectively penetrate the clumps. This explains the deficit of CRs with energies below $\approx 10$~TeV in the clumps. The very hard spectral slope found below the peak is an effect of the steep energy dependence of the Bohm diffusion coefficient. The position of the peak moves towards larger energies for clumps that enter later the SNR shock, as it can be inferred by Eq.~\ref{eq:difftime} and the discussion that follows it.

\begin{figure}
\centering
\includegraphics[width=.39\textwidth]{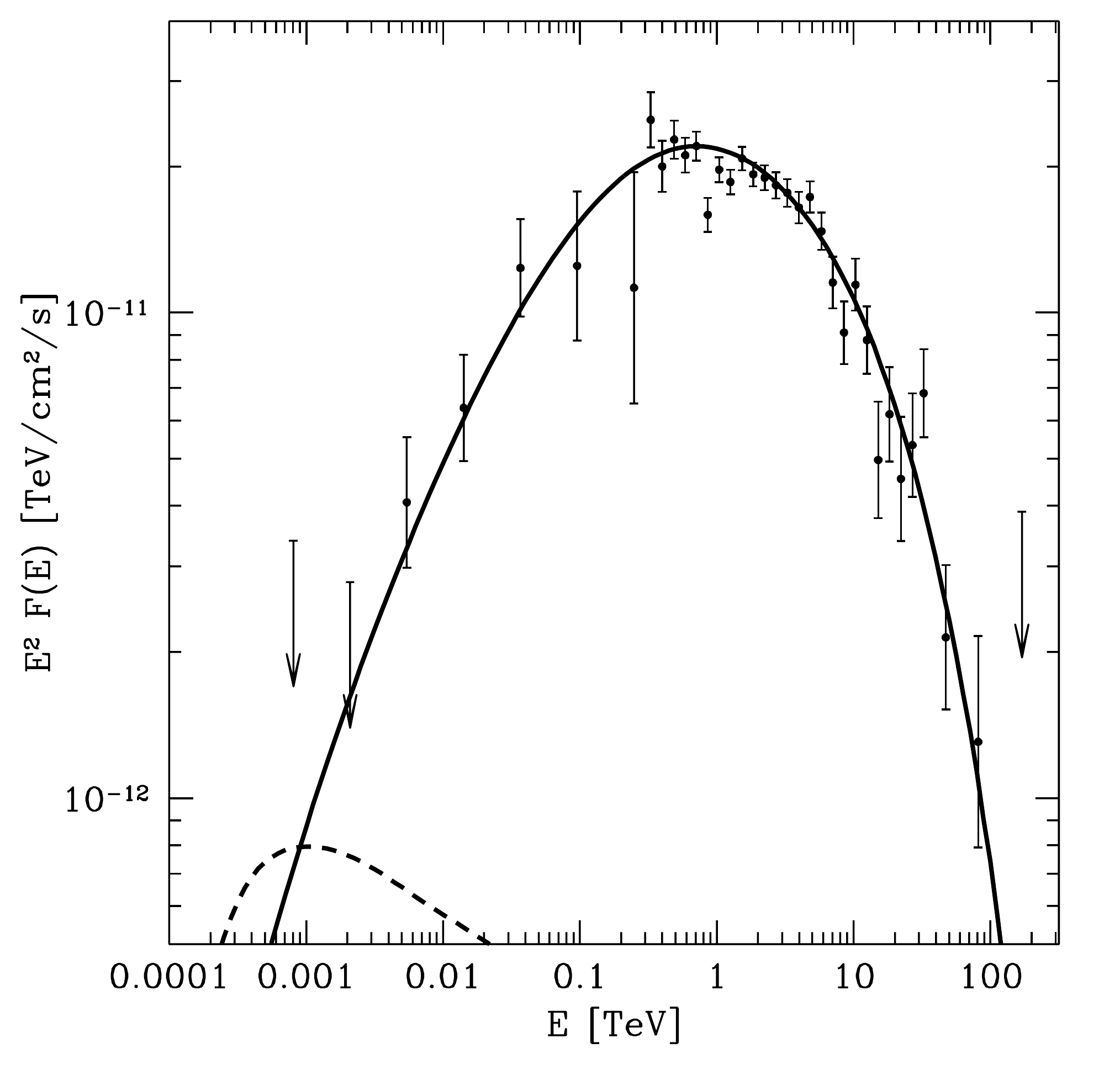}
\caption{Gamma--rays from RX~J1713.7-3946. The emission from the clumps is shown as a solid line, while the dashed line refers to the emission from the diffuse gas in the shell. Data points refer to {\it FERMI} and {\it HESS} observations.}
\label{fig2}
\end{figure}

The hadronic gamma--ray emission from all the dense clumps in the shell is plotted as a solid line in Fig.~\ref{fig2}. 
The gas density within clumps is $n_{c,3} = 1$ and the density of clumps is 3~pc$^{-3}$, which implies a total mass in the clumps within the SNR shell of $550~M_{\odot}$ and a clump volume filling factor of $\approx 0.01$. The distance to the SNR is 1~kpc. The prediction is in agreement with {\it FERMI} and {\it HESS} data. The gamma--ray emission from CR interactions in the low density diffuse gas swept up by the SNR is plotted as a dashed line, and shown to be subdominant. The contribution from inverse Compton scattering from electrons accelerated at the SNR is expected to be negligible, if the magnetic filed is $\gtrsim 10~\mu$G.

Secondary electrons are also produced in proton--proton interactions in the dense clumps. Their production spectrum is similar in shape to that of gamma--rays (Fig.~\ref{fig2}), with a normalization smaller by a factor of $\approx 2$ and particle energies larger by the same factor. Thus, the peak of electron production happens at an energy of $\approx 2$~TeV. Such electrons escape the clump in a time $\sim 200$~yr (Eq.~\ref{eq:difftime}), which is shorter than both synchrotron and Bremmstrahlung energy loss time ($\sim 450$ and $3.3 \times 10^4$~yr, respectively, for the parameters considered here). Thus, no contribution from secondary electrons has to be expected to the gamma--ray emission.

At the present time, the CRs inside RX~J1713.7-3946 amounts to $\approx 6~\%$ of the total explosion energy. However, the fraction of explosion energy converted into CRs is likely to increase with time, given that the SNR is still in a quite early stage of its dynamical evolution. It can be shown that the total fraction of explosion energy converted into CRs by a SNR is of the same order of the istantaneous CR acceleration efficiency $\eta$ \citep[e.g.][]{meescape}. The value $\eta = 0.1$ adopted here is consistent with the typical acceleration efficiency that SNRs should have in order to be the sources of Galactic CRs.

\section{Discussion and conclusions}

We have shown that the gamma--ray emission from RX~J1713.7-3946 can be naturally explained by the decay of neutral pions produced in hadronic CR interactions with a dense, clumpy gas embedded in the SNR shell.
The clumps are expected to be surrounded by a turbulent layer characterized by an average magnetic field of $\approx 100~\mu$G, which in some cases can reach values as large as $\approx 1$~mG \citep{inoue2012}.
Such extreme values would explain the very rapid (year--scale) time variability observed in X--rays from small knots inside the SNR shell \citep{yas}.

Also the absence of thermal X--ray emission from the SNR fits well within this scenario. This is because the SNR shock propagates in the low density, inter--clump medium, and not in the dense clumps, which remain cold and unable to emit thermal X--rays. By means of a numerical simulation, \citet{don} estimated the thermal X--ray emission expected for RX~J1713.7-3946 for a value of the ambient density of $n_{h,-2} = 5$ (a factor 2.5 larger than the one adopted here). They found that the expected emission is subdominant with respect to the X--ray synchrotron emission from the SNR, as it is indeed observed \citep{tanaka}.

As noticed in the introduction, multi--zone leptonic models might also provide a satisfactory fit to the observed multi--wavelength spectrum of RX~J1713.7-3946. Thus, further observational evidences are needed in order to discriminate between the hadronic and leptonic origin of the gamma-ray emission. A conclusive proof of the validity of the hadronic scenario would come from the detection of neutrinos. This test is feasible, since in this scenario the expected neutrino flux from RX~J1713.7-3946  is within the reach of km$^3$--scale detectors \citep[e.g.][]{francesco}. 

\section*{Acknowledgments}

We thank S. Casanova, Y. Fukui, S. Orlando, Y. Uchiyama, J. Vink, and V. Zirakashvili for helpful discussions.
SG acknowledges support from the UnivEarthS Labex program at Sorbonne Paris Cit\'e (ANR-10-LABX-0023/ANR-11-IDEX-0005-02), and API and GRAPPA Institutes at the University of Amsterdam for kind hospitality.

\label{lastpage}

\end{document}